\documentclass[11pt,letterpaper]{article}
\usepackage[margin=1.2in]{geometry}
\usepackage[blocks, affil-it]{authblk}
\usepackage{amsmath}
\usepackage{amssymb}
\usepackage{array}
\usepackage{listings}
\usepackage{empheq}
\usepackage{amsthm}
\usepackage{eucal}
\usepackage{ulem}
\usepackage{bbm}
\usepackage{titlesec}
\usepackage{fullpage}
\usepackage{enumerate}
\usepackage{fancyhdr}
\usepackage{graphicx}
\usepackage{subcaption}
\usepackage{booktabs,tabularx,colortbl} 
\usepackage{xcolor}
\usepackage{algpseudocode}
\usepackage{algorithm}
\usepackage{dsfont}
\usepackage[utf8]{inputenc}
\usepackage[pdfencoding=auto, psdextra]{hyperref}
\hypersetup{
    colorlinks=true,
    linkcolor=blue,
    filecolor=magenta,      
    urlcolor=blue,
    citecolor=blue
}
\usepackage{fullpage, lastpage, fancyhdr}
\usepackage{tocloft}
\usepackage{natbib}
\usepackage{thmtools}

\usepackage{amsfonts}
\usepackage{algpseudocode}
\usepackage{algorithm}
\usepackage{tikz}
\usepackage{titlesec}
\usepackage{fullpage}
\usepackage{lastpage}
\DeclareMathOperator*{\argmin}{argmin}

\renewcommand\P{\mathbbm{P}}
\newcommand\E{\mathbbm{E}}

\newcommand{\lfdr}{\text{lfdr}}

\newcommand{\bFDR}{\textnormal{bFDR}}
\newcommand{\FDR}{\textnormal{FDR}}
\newcommand{\FWER}{\textnormal{FWER}}
\newcommand{\SL}{\textnormal{SL}}
\newcommand{\BH}{\textnormal{BH}}
\newcommand{\MASL}{\textnormal{MASL}}
\newcommand{\MABH}{\textnormal{MABH}}

\newcommand{\tol}{\alpha}

\newtheorem{theorem}{Theorem}

\newtheorem{corollary}[theorem]{Corollary}
\newtheorem{proposition}[theorem]{Proposition}

% ...

\title{A minimally adaptive Support Line procedure}

\author[1]{Sarah Mostow}
\author[2]{Nikolaos Ignatiadis}
\author[2]{Daniel Xiang}
\affil[1]{Duke University}
\affil[2]{University of Chicago}

\begin{document}

\maketitle

\begin{abstract}
The Support Line (SL) method is a multiple testing procedure introduced by \cite*{soloff2024edge} for controlling the boundary false discovery rate (bFDR), defined as the type~I error probability for the discovery made at the boundary of the rejection region. We propose an adaptive variant of SL, called Minimally Adaptive Support Line (MASL), that improves the power of SL uniformly while controlling the bFDR under independence. Numerically, this power gain is largest when the number of hypotheses is small. We study the case of $m=2$ hypotheses in detail, which allows us to contrast bFDR, FDR, and family-wise error rate (FWER) control. 
In this setting, MASL is shown to control bFDR without also controlling the FWER, whereas SL controls both bFDR and FWER.
We also use this case to illustrate why a lack of monotonicity is natural for bFDR-controlling procedures such as SL and MASL.
\end{abstract}


\section{Introduction}

In multiple testing contexts, it is desirable to make decisions that are individually justified. A natural approach appeals to a monotonicity heuristic, namely, that a smaller $p$-value indicates more evidence against its null hypothesis. Formally, suppose the analyst observes $p$-values $p_1,\dots,p_m$ corresponding to null hypotheses $H_1,\dots,H_m$, and rejects those for which $p_i \leq t$. 
The least compelling discovery would then correspond to the rejected null hypothesis with the largest $p$-value. If the procedure ensures this boundary rejection is defensible, then by the monotonicity heuristic, so are the rest. One formalization of this criterion is the boundary false discovery rate (bFDR, \citealp*{xiang2026frequentist}): for a procedure rejecting the null hypotheses with the $R$ smallest $p$-values, the boundary FDR is the probability that the last rejection is a false discovery,
\begin{align*}
    \bFDR(R) = \P(H_{(R)} \text{ is true}),
\end{align*}
where $H_{(1)},\dots,H_{(m)}$ are the null hypotheses ordered according to $p_{(1)}\leq \dots \leq p_{(m)}$, taking the convention $H_{(0)}\coloneqq \text{false}$.
While $H_1,\dots,H_m$ are viewed here as having fixed (non-random) truth statuses, $\P(H_{(R)}\text{ is true})$ is generally a non-trivial probability since the index of the boundary rejection depends on the data. 

The false discovery rate (FDR, \citealp{benjamini1995controlling}) is a property of multiple testing procedures that measures the quality of their discoveries in an average-case sense. It is defined in terms of a uniformly selected rejection $\FDR = \P(H_{(I)} \text{ is true})$ where $I$ is a uniform draw from $\{1,\dots,R\}$ and we set $I\coloneqq 0$ on the event $\{R=0\}$. The bFDR is a local version that focuses on the boundary rejection, and is usually larger because less promising rejections $(p\approx t)$ are aggregated with strong bets ($p\approx 0$) when computing FDR.

When the $p$-values are jointly independent with Uniform$(0,1)$ null distribution, the Support Line (SL) procedure of \cite*{soloff2024edge} controls the bFDR at level $\pi_0 \tol$, where $\pi_0$ is the proportion of true null hypotheses and $\alpha \geq 0$ is a pre-specified type-I error tolerance. 

The SL$(\alpha)$ procedure rejects the $R_\tol$ smallest $p$-values, where
\begin{align*}
    R_\tol = \argmin_{k=0,\dots,m} \left\{ p_{(k)}-\frac{\tol k}{m} \right\}, \hspace{2em} p_{(0)} \coloneqq 0,
\end{align*}
using the convention that $R_\alpha$ is the largest element of the $\argmin$ if the minimizer is not unique.

Procedures that improve power by adapting to the number of true null hypotheses while maintaining bFDR control are referred to as \textit{adaptive} bFDR procedures. While these generally achieve bFDR closer to $\alpha$ with higher power than SL, they may underperform when $\pi_0$ is close to $1$ \citep{mostow2026adaptive}.

In this note, we present one such procedure that always makes at least as many rejections as SL$(\alpha)$, while maintaining $\bFDR \leq \alpha$ for any configuration of $H_1,\dots,H_m$. The method plays an analogous role to the minimally adaptive Benjamini-Hochberg procedure (MABH, \cite{solari2017minimally})

as well as its $e$-value variant \citep[Supplement 3.2]{ignatiadis2024evalues} that provide small but uniform improvements to BH \citep{benjamini1995controlling} and $e$-BH \citep{wang2022false} while controlling FDR under the exact same conditions as the original procedures. 
The principle underlying these procedures is similar to stepping-down arguments for control of the family-wise error rate (see, e.g.,~\cite{goeman2010sequential}).\footnote{\citet{ignatiadis2026tiny} uniformly improve upon adaptive BH procedures using a different principle, that of compound e-values~\citep{ignatiadis2025asymptotic}.
}

\begin{figure}[t]
    \centering
    \includegraphics[width=0.8\linewidth]{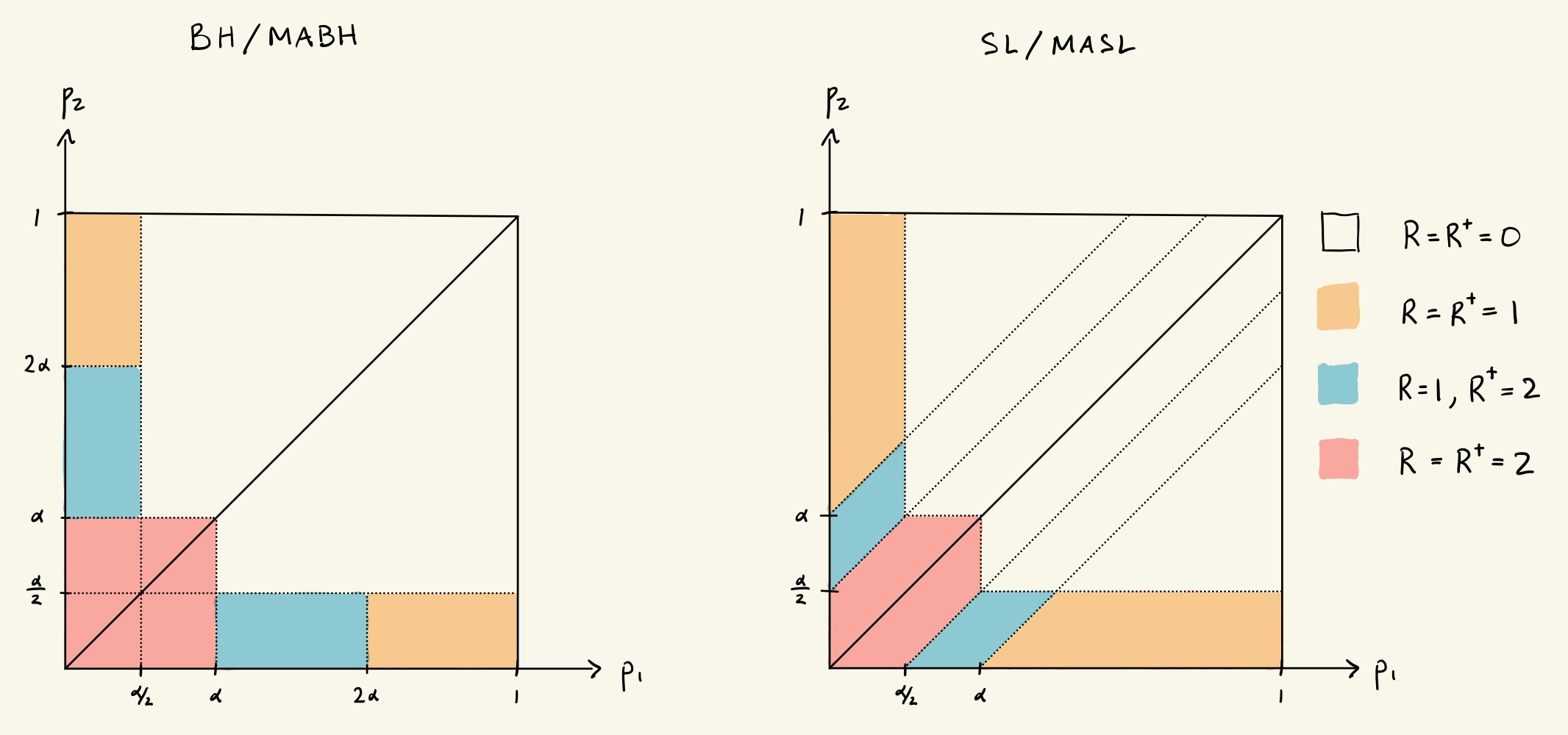}
    \caption{For $m=2$, the above figure shows the set of $(p_1,p_2)$ for which SL and MASL (right) make 0, 1 or 2 rejections. The analogous picture for BH and MABH is shown on the left (see also Figure 1 in \cite{solari2017minimally}). The non-shaded region means 0 rejections by both procedures.}
    \label{fig:MASL-Rej}
\end{figure}

\section{Main result}

The minimally adaptive support line procedure (MASL) is defined as follows for $\alpha \geq 0$. If there is just $m=1$ hypothesis, then a level $\alpha$ test is performed for this hypothesis. 
Otherwise, $m\geq 2$ and the MASL$(\alpha)$ procedure rejects the $R_\tol^+$ smallest $p$-values, where
\begin{align*}
    R_\tol^+ = \begin{cases}
        0 \hspace{1em} &\text{if } R_\tol=0 \\
        \argmin_{k=0,\dots,m} \left\{ p_{(k)}-\frac{\tol k}{m-1} \right\} &\text{if } R_\tol>0.
    \end{cases}
\end{align*}
MASL$(\alpha)$ is equivalent to SL$(\alpha/\hat{\pi}_0)$ where $\hat{\pi}_0 = \frac{m-1}{m}$ on the event that some null hypothesis is rejected by SL$(\alpha)$ and $\hat{\pi}_0=1$ otherwise.
When $R_\tol=0$, MASL also makes no rejections. When $R_\tol>0$, MASL makes at least as many rejections as SL. Figure \ref{fig:MASL-Rej} shows the rejection regions of the two procedures in the special case of $m=2$ hypotheses. 
For $\alpha\leq 1/2$, the area of the region in which MASL makes strictly more rejections than SL is $\tol^2/2$ (half the improvement of MABH over BH in the same setting). 
The following proposition is our main result. 
\begin{proposition}
\label{prop:main}
    If $\bFDR(R_\tol)\leq \pi_0 \tol$ for all $\tol \geq 0$, then $\bFDR(R^+_\tol) \leq \tol$ for all $\tol \geq 0$.
\end{proposition}

\begin{proof}
    Suppose $\pi_0=1$. Then $\bFDR(R_\tol) = \P(R_\tol >0) = \P(R_\tol^+>0) = \bFDR(R_\tol^+)$ by definition of $R_\tol^+$. Now suppose $\pi_0<1$. If $R_\tol=0$, then $R_\tol^+=0$ and so $1\{H_{(R_\tol)}\text{ is true}\} = 1\{H_{(R_\tol^+)}\text{ is true}\} = 0$. If $R_\tol>0$, then $R_\tol^+ = R_{m\tol/(m-1)}$ so that $1\{H_{(R^+_\tol)}\text{ is true}\}=1\{H_{(R_{m\alpha /(m-1)})}\text{ is true}\}$.
    Together this gives $1\{H_{(R^+_\tol)}\text{ is true}\}\leq 1\{H_{(R_{m\alpha /(m-1)})}\text{ is true}\}$ almost surely, so that $\bFDR(R^+_\tol) \leq \bFDR(R_{m\tol/(m-1)}) \leq \frac{\pi_0  m\tol}{m-1} \leq \tol$.
\end{proof}
\begin{corollary}
\label{cor:MASL-bFDR-bound}
    If $p_1,\dots,p_m$ are independent and $p_i \sim \text{Uniform}(0,1)$ when $H_i$ is true, then for $\alpha \geq 0$,
    \begin{align*}
        \bFDR(R^+_\alpha) \leq \alpha.
    \end{align*}
\end{corollary}
In practice, the improvement of MASL over SL is appreciable only when the number of hypotheses is small. The special case with $m=2$ hypotheses is illustrated numerically in the next section.

\section{Elaboration for $m=2$ hypotheses}

Since the adaptive procedures' level $\{m/(m-1)\}\alpha$ is largest when $m=2$, we give special attention to this case. Figure~\ref{fig:MASL-Rej} shows the rejection regions for $m=2$, from which we see that
\begin{equation}
\label{eq:ordering}
\MABH \succ \MASL \succ \BH \succ \SL,
\end{equation}
where $\mathcal{B} \succ \mathcal{A}$ means that procedure $\mathcal{B}$ makes at least as many rejections as procedure $\mathcal{A}$ for every $(p_1, p_2) \in [0,1]^2$, and strictly more rejections on a set of positive Lebesgue measure.
The gain of MASL over BH is small, and these two procedures only differ on two triangles with combined area $\alpha^2/4$.

\subsection{Comparing type-I error metrics and different methods}
\label{subsec:type-1-comparison}

\begin{table}[tbp]
  \centering
  \caption{Error-rate control of the four procedures for $m=2$
    independent $p$-values with $\mathrm{Uniform}(0,1)$ null distributions.
   Rejections are nested as in~\eqref{eq:ordering} with
$\SL \succ \BH \succ \MASL \succ \MABH$.}
  \label{tab:procedure-error-control}

  \begingroup
  \definecolor{maslControl}{HTML}{E3EFEB}
  \definecolor{maslFailure}{HTML}{F6E5DA}
  \newcommand{\maslControlCell}[1]{%
    \cellcolor{maslControl}#1}
  \newcommand{\maslFailureCell}{%
    \cellcolor{maslFailure}Can exceed $\alpha$}
  \newcommand{\maslRowHeight}{\rule[-1.1ex]{0pt}{4.2ex}}
  \setlength{\tabcolsep}{3pt}
  \setlength{\extrarowheight}{0pt}
  \renewcommand{\arraystretch}{1}

  \vspace{0.7ex}
  \begin{tabularx}{\linewidth}{@{}l*{4}{>{\centering\arraybackslash}X}}
    \toprule[0.7pt]
    & \multicolumn{4}{c}{Procedure} \\
    \cmidrule(lr){2-5}
    Error rate & MABH & MASL & BH/Hochberg & SL \\
    \midrule[0.4pt]
    \maslRowHeight FWER
      & \maslFailureCell
      & \maslFailureCell
      & \maslControlCell{$\leq\alpha$}
      & \maslControlCell{$\leq\alpha$} \\
    \addlinespace[3pt]
    \maslRowHeight bFDR
      & \maslFailureCell
      & \maslControlCell{$\leq\alpha$}
      & \maslControlCell{$\leq\alpha$}
      & \maslControlCell{$=\pi_0\alpha$} \\
    \addlinespace[3pt]
    \maslRowHeight FDR
      & \maslControlCell{$\leq\alpha$}
      & \maslControlCell{$\leq\alpha$}
      & \maslControlCell{$=\pi_0\alpha$}
      & \maslControlCell{$\leq\alpha$} \\
    \bottomrule[0.7pt]
  \end{tabularx}

  \endgroup
\end{table}

We find it instructive to elaborate on the ordering in~\eqref{eq:ordering} by also summarizing what error metrics are controlled by each procedure for $m=2$ hypotheses with independent $p$-values and uniform null distribution. Table~\ref{tab:procedure-error-control} summarizes these properties; a proof for each entry in the table is recorded in Appendix~\ref{sec:proofs-table-1}.
First, as noted by~\cite{solari2017minimally}, BH is equivalent to Hochberg's procedure~\citep{hochberg1988sharper} for $m=2$ and so controls the FWER. Consequently, BH also controls bFDR at level $\alpha$ (but not at $\pi_0\alpha$; see Appendix~\ref{sec:proofs-table-1}). 
The same is not true for MABH, whose bFDR equals $2\alpha$ if $m_0=1$, the non-null $p$-value is exactly zero and $\alpha \in (0,1/2)$. Thus MABH controls the FDR but neither the bFDR nor the FWER (\citet{solari2017minimally} already noted that MABH does not control FWER).
Meanwhile, MASL controls the bFDR (and FDR) without also controlling the FWER.

The ordering in~\eqref{eq:ordering} along with the equivalence of BH and Hochberg for $m=2$ also has an implication beyond independence: SL controls both FWER and bFDR under PRDS if the null distribution is super-uniform (Proposition~\ref{prop:prds} in Appendix~\ref{subsec:hochberg}). By contrast, MASL controls neither of these error criteria under PRDS (see Appendix~\ref{sec:proofs-table-1} and Figure~\ref{fig:bfdr-SL-MASL}).

\subsection{Nonmonotonicity of SL and MASL}
\label{subsec:monotonicity}

\begin{figure}
    \centering
    \includegraphics[width=0.7\linewidth]{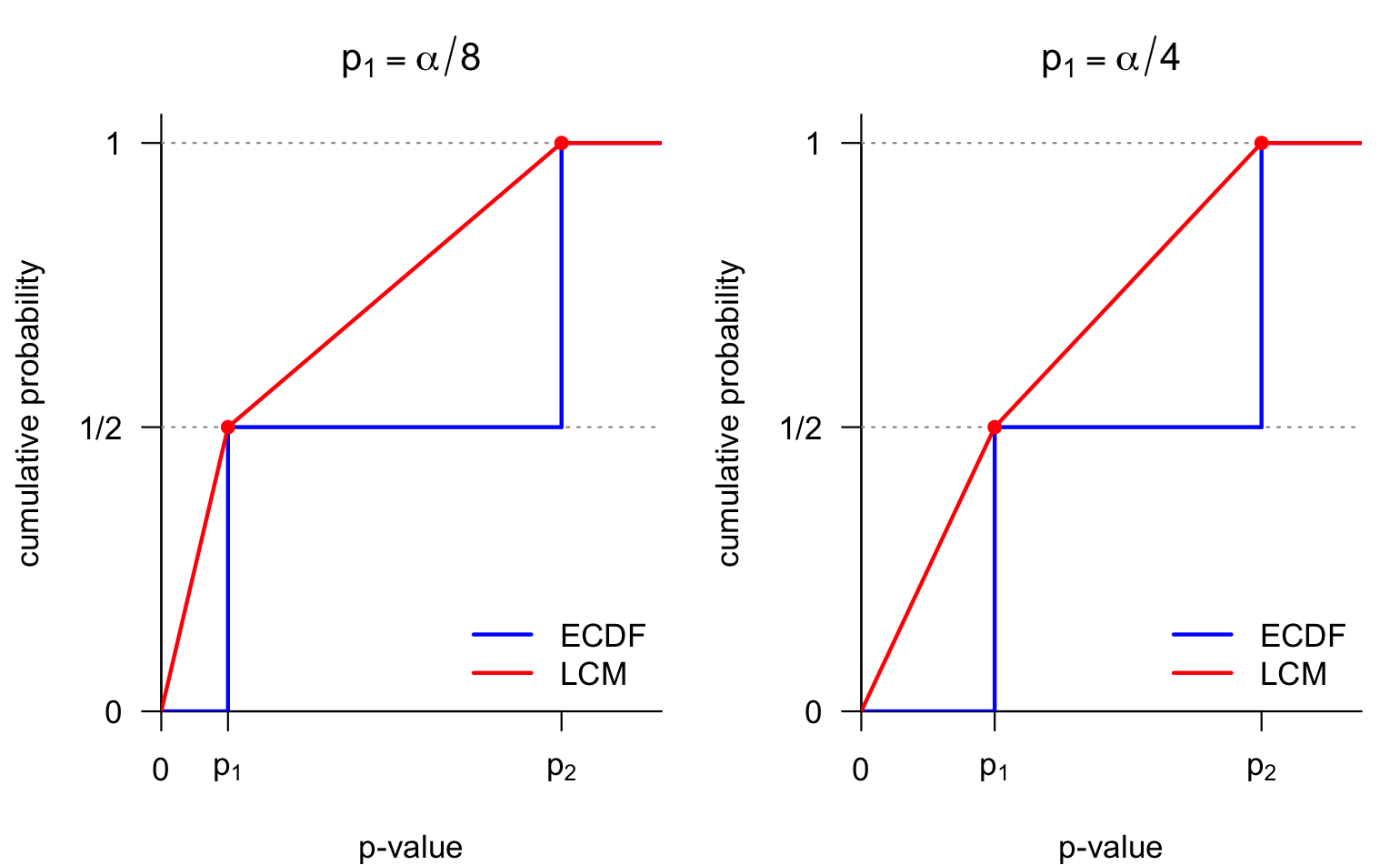}
    \caption{Empirical CDF (ECDF) and Least Concave Majorant (LCM) for $p_2 = 3 \alpha / 4$ (fixed) and $p_1 = \alpha / 8$ (left) vs $\alpha / 4$ (right), with $\alpha = 0.2$. Increasing $p_1$ from $\alpha /8$ to $\alpha / 4$ decreases $\hat f (p_1)$ from $20$ to $10$ and increases $\hat f (p_2)$ from $4$ to $5$.}
    \label{fig:nonmononotonicity}
\end{figure}

Due to the regions in the right panel of Figure \ref{fig:MASL-Rej} having slanted borders, the SL and MASL procedures are not monotone, in the sense that increasing a $p$-value can increase the number of rejections. For example, $(p_1,p_2) = (\alpha/8,3\alpha/4)$ leads to one rejection by SL, but if $p_1$ is increased to $\alpha/2$ while holding $p_2$ fixed, then SL makes two rejections. This non-monotonicity arises because SL and MASL compare the spacing $p_{(2)}-p_{(1)}$ to a threshold in order to decide how many hypotheses to reject,\footnote{By contrast, BH compares each $p_{(k)}$ to the fixed threshold $k\alpha/m$, and MABH to $k\alpha/(m-1)$ whenever BH makes at least one rejection.} and may be interpreted through the connection between SL and the Grenander estimator for a monotone density \citep{grenander1956theory,soloff2024edge}; also see~\citet{wadekar2026calibration}. SL rejects hypothesis $H_{(k)}$ when the Grenander density estimator at $p_{(k)}$ exceeds $1/\alpha$, i.e. its estimated $\widehat{\lfdr}=1/\hat{f}_m(p_{(k)})$ is at most $\alpha$. If $p_{(1)}$ is brought closer to $p_{(2)}$, then the density estimate at $p_{(2)}$ becomes larger, thus decreasing its lfdr estimate, and there is stronger evidence to reject. This is illustrated in Figure~\ref{fig:nonmononotonicity} with $\alpha = 0.2$, where increasing $p_1$ from $\alpha /8$ to $\alpha / 4$ increases $\widehat{\lfdr} (p_1)$ from $0.05$ to $0.10$ and decreases $\widehat{\lfdr} (p_2)$ from $0.25$ to $0.2$.

A similar phenomenon arises with fixed cutoffs, and is tied to the bFDR criterion itself. To illustrate, consider the thresholding procedure $\mathcal{T}_s$ that rejects all hypotheses with $p_i \leq s$. Suppose $p_1 \sim \mathrm{Uniform}(0,1)$ is null and $p_2 \sim \mathrm{Uniform}(0,\tau)$ for $\tau \in (0,1)$ is the non-null $p$-value and that both $p$-values are independent. For $\tau \leq s$, $\mathrm{bFDR}(\mathcal{T}_s) = s - \tau/2$. Thus, for $\tau \le 2\alpha$ the threshold $s$ achieving $\mathrm{bFDR}(\mathcal{T}_s) = \alpha$ is $s = \alpha + \tau/2$. (By contrast, for $\tau \leq s$, $\mathrm{FDR}(\mathcal{T}_s) = s/2$ does not depend on $\tau$.) As the non-null $p$-value becomes stochastically smaller (i.e., as $\tau$ becomes smaller), $p_2$ is rejected with certainty and pushed further toward zero, so whenever the null is rejected it is increasingly likely to lie above $p_2$ and hence to be the last discovery. Keeping the bFDR at $\alpha$ therefore requires a smaller threshold, even though the signal has become stronger. Since $\tau$ is unknown, SL and MASL must in effect implement a data-driven version of this threshold, which is the source of their non-monotonicity.

This non-monotonicity property stands in direct contrast to the frequently criticized “free-rider” phenomenon in FDR-controlling procedures such as BH~\citep{finner2001false}. Under these procedures, strengthening the signals of a small number of hypotheses can substantially reduce the overall FDR of a potential rejection set, thereby allowing hypotheses with considerably weaker evidence to enter the rejection set as “free-riders” or “hitchhikers,” despite their limited individual evidence against the null. However, SL exhibits the opposite behavior; as the evidence against a subset of hypotheses strengthens, larger p-values whose evidence appears weaker by comparison become less likely to be rejected.

Finally, we note that \citet{ramdas2026minimally} recently showed that MABH is admissible among monotone FDR-controlling procedures. While monotonicity is a natural restriction for FDR control~\citep{blanchard2008two}, our discussion in this section suggests that it is not appropriate for bFDR control. Whether MASL enjoys an analogous admissibility property within a suitable class of (non-monotone) bFDR-controlling procedures remains an open question.

\subsection{Numerical results}

\begin{figure}[h]
    \centering
    \includegraphics[width=\linewidth]{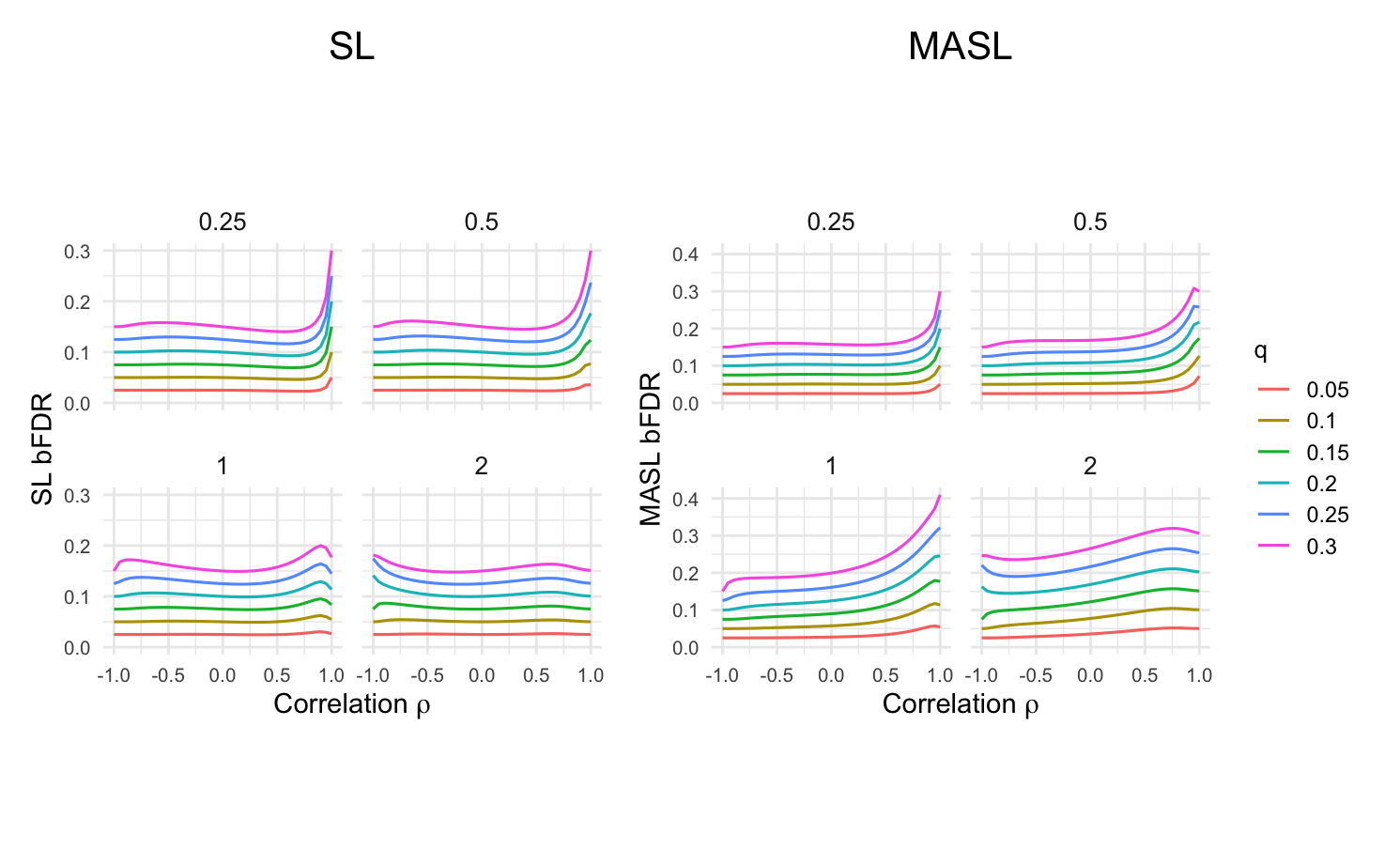}
    \caption{bFDR (vertical axis) versus correlation (horizontal axis) for SL and MASL run on $m=2$ $p$-values arising from one-sided Gaussian location testing with correlation $\rho$, $H_1$ is true, $H_2$ is false, and $\mu \in \{0.25,0.5,1,2\}$ in the model \eqref{eq:one-sided-tests}.}
    \label{fig:bfdr-SL-MASL}
\end{figure}

We examine both SL and MASL in a one-sided location testing problem with $m=2$ test statistics, 
\begin{align}
\label{eq:one-sided-tests}
    (z_1, z_2) \sim N \left( \begin{pmatrix} 0 \\ \mu \end{pmatrix}, \begin{pmatrix} 1 & \rho \\ \rho & 1 \end{pmatrix}\right),
\end{align}
where $\mu \in \mathbb{R}$ and $\rho \in [0,1]$. The $p$-values are computed as $p_i=1-\Phi(z_i)$ for $i=1,2$, where $\Phi$ is the cdf of the $N(0,1)$ distribution.
For $\mu = 0$, i.e.,~the global null ($\pi_0 = 1$), we observe the following three facts. Both SL and MASL have equal bFDR since $\{R_\tol>0\}=\{R_\tol^+>0\}$, SL makes at least one discovery if and only if BH makes at least one discovery, and finally, $\FDR=\bFDR=\FWER$.
Consequently, under the global null, both SL and MASL control type 1 error under general forms of positive dependence \citep{sarkar1998some,benjamini2001control} that include model~\eqref{eq:one-sided-tests} for any $\rho \in [0,1]$.
For the $\pi_0 = 1/2$ case, we calculated bFDR for $\mu = 0.25, 0.5, 1, 2$. Figure \ref{fig:bfdr-SL-MASL} shows that SL controls bFDR at or below level the nominal level $\tol$ while MASL does not, particularly for $\mu = 1$ and $\mu = 2$. A proof for the former claim is provided in Appendix \ref{subsec:hochberg}.

\begin{figure}[h]
    \centering
    \includegraphics[width=1\linewidth]{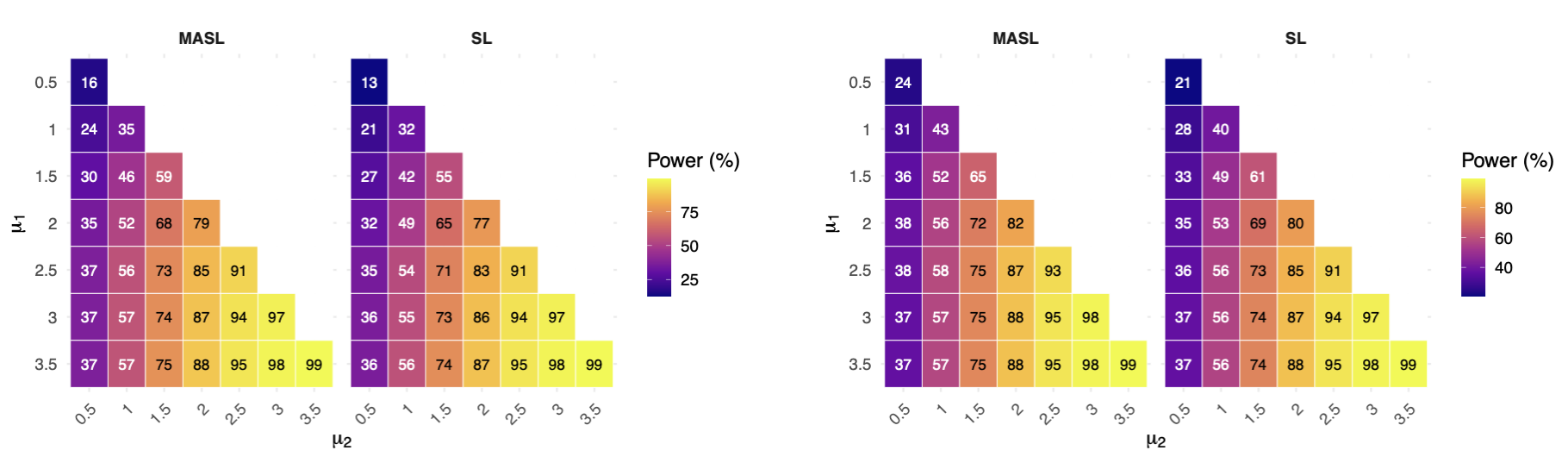}
    \caption{A power comparison between SL and MASL for $m=2$ hypotheses. The heatmaps show $\P(\text{both hypotheses rejected}$) when both $H_1,H_2$ are false null hypotheses,  
    where $p_1,p_2$ are one-sided $p$-values computed from a Gaussian bivariate model with correlation $\rho =0$ (left) or $\rho=0.5$ (right), and means $\mu_1,\mu_2$ varying between 0.5 and 3.5.}
    \label{fig:one-sided-Gaussian-expt}
\end{figure}

\begin{figure}[h!]
    \centering
    \includegraphics[width=0.75\linewidth]{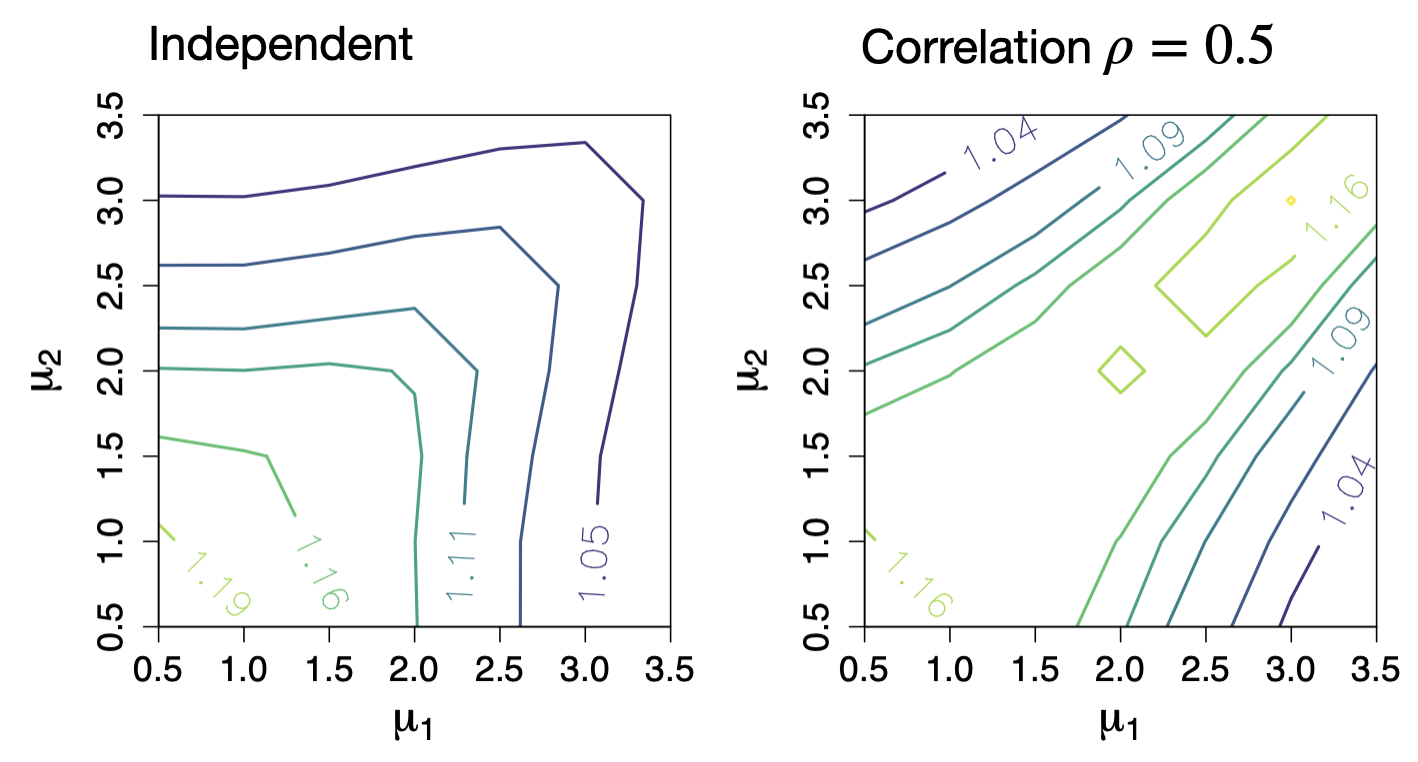}
    \caption{Contour plot of the power odds ratio between MASL and SL as in the setting of Figure~\ref{fig:one-sided-Gaussian-expt}.}
    \label{fig:contour}
\end{figure}

Figures \ref{fig:one-sided-Gaussian-expt} and \ref{fig:contour} show power comparisons in the case where both null hypotheses are false, i.e.
\begin{align*}
    (z_1, z_2) \sim N \left( \begin{pmatrix} \mu_1 \\ \mu_2 \end{pmatrix}, \begin{pmatrix} 1 & \rho \\ \rho & 1 \end{pmatrix}\right).
\end{align*}
when $\rho=0$ and $\rho=0.5$, where the gains are typically between 0 and 3\% as $\mu_1,\mu_2$ vary between $0.5$ and $3.5$.

\section{Further numerical results}

\begin{figure}[htbp]
  \centering
  \begin{subfigure}[b]{0.48\textwidth}
    \centering
    \includegraphics[width=\textwidth]{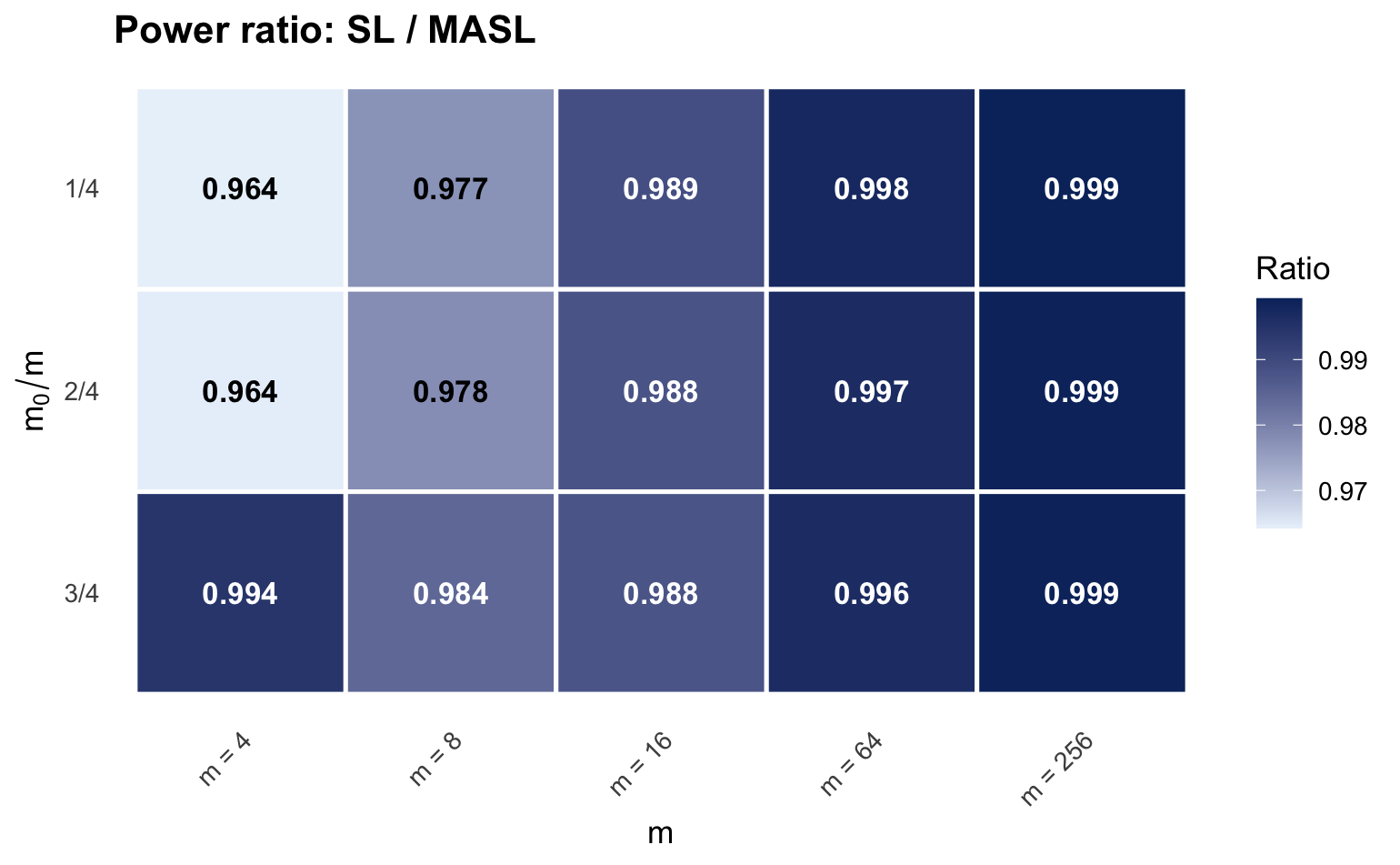}
    \caption{Relative power of SL to MASL over $N=10,000$ simulations.}
    \label{fig:power_sl_vs_masl}
  \end{subfigure}
  \hfill
  \begin{subfigure}[b]{0.48\textwidth}
    \centering
    \includegraphics[width=\textwidth]{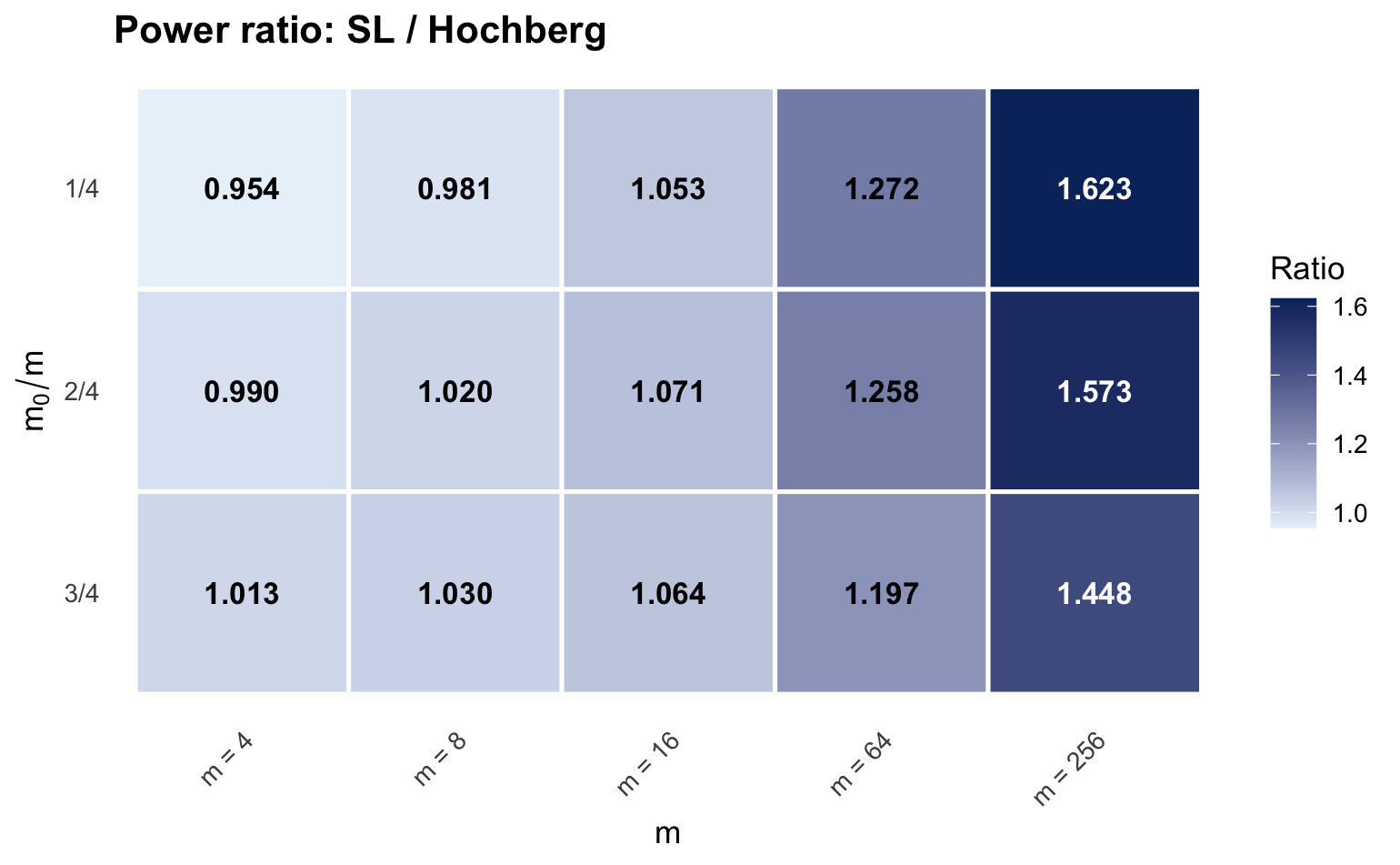}
    \caption{Relative power of SL vs Hochberg over $N=~10,000$ simulations}
    \label{fig:power_sl_vs_hoch}
  \end{subfigure}
  \caption{Relative power of procedures across different values of $m$ and $m_0/m$.}
  \label{fig:power_comparison}
\end{figure} 

To explore the relative performances of SL, MASL, and Hochberg for larger numbers of hypotheses, we generated $m \in \{4,8,16,64,256\}$ one-sided $p$-values of independent unit-variance Gaussian test statistics with means as follows: $m_0$ ``null" means were set to 0. For the remaining $m - m_0$ $p$-values, means were drawn uniformly from $\{1, 2, 3, 4\}$. SL, MASL, and Hochberg procedures were run at tolerance level $\alpha = 0.2$, and power (the proportion of non-nulls rejected) was recorded for each procedure. Figure~\ref{fig:power_comparison} illustrates the power ratios of SL vs MASL and SL vs Hochberg. As expected, SL always has lower power than MASL, and the ratio of SL power to MASL power is closer to 1 for larger values of $m$. The heatmap of SL power vs Hochberg power, however, indicates that neither SL nor Hochberg uniformly dominates the other procedure for all values of $m$.

\section{Discussion}

In this note, we demonstrated that the Support Line procedure can be uniformly improved by an adaptive modification, regardless of the true value of $\pi_0$. 

The scope of validity for MASL is currently only established for the case with continuous and independent $p$-values with Uniform$(0,1)$ null distribution, where
the original SL procedure controls bFDR at $\pi_0 \alpha$. For the case of $m=2$ hypotheses, Proposition \ref{prop:prds} shows that SL controls bFDR below $\alpha$ under positive dependence, but this does not imply the condition of Proposition \ref{prop:main}; indeed, Figure \ref{fig:bfdr-SL-MASL} exhibits a failure of MASL to control bFDR in that setting.
Less restrictive conditions under which SL maintains the bound at $\pi_0 \alpha$ are not yet well understood. Some progress has been made relaxing these assumptions to settings with dependent and discrete $p$-values \citep{gao2026conformal}, but a sharper picture has yet to emerge.

The idea of minimally adaptive procedures has a longer history. An early
example is Scheff\'e's ``modified S-method''~\citep{scheffe1970multiple}, which
applies an improved Scheff\'e procedure after the global null
hypothesis has been rejected. The same idea is also closely tied to FWER stepping-down arguments. It is natural to ask whether a multi-stage extension of the MASL procedure controls bFDR, in which the estimate of $\pi_0$ decrements as more rejections are made. However, the logic underlying the sequential rejection principle for FWER control \citep{goeman2010sequential} does not readily apply to the bFDR criterion. Once FWER control is lost on a smaller discovery set, the analyst can proceed aggressively to reject more hypotheses, but the analogous statement does not hold for bFDR; in this case, the type-I error event on a subset of discoveries does not imply the error event on the full set.

Recent work by \cite{goeman2026uniform}
proposed a new family of methods that uniformly improve upon BH using the e-closure principle~\citep{xu2025bringing}, a generalization of the classical closure principle for FWER control~\citep{marcus1976closed}. Unlike MABH, the new procedures can provide substantial power gains even when $m$ is large. Whether there is an analogous extension for SL using the e-closure principle is an ongoing investigation by the authors.

\paragraph{Acknowledgments.} We thank the Editor and two anonymous referees for helpful comments, and in particular the referee who suggested the proof of Proposition \ref{prop:prds} and the connection to Scheff\'e's modified S-method.

\paragraph{Declaration of generative AI use.} An AI model (Claude, Fable 5.2) was used to help draft Appendix~\ref{sec:proofs-table-1}. Claude (Opus 5.5) and ChatGPT (GPT-6) were also utilized for some copyediting feedback and to write parts of the code for the figures and the table in the paper.

\bibliographystyle{dcu}
\bibliography{reference}


\appendix
\section{Appendix}
\label{sec:appendix}

\subsection{Hochberg's procedure and error-control under PRDS}
\label{subsec:hochberg}

The procedure of \cite{hochberg1988sharper} is recalled here: for any $i=m,m-1,\dots,1$, if
\begin{align*}
    p_{(i)} \leq \frac{\alpha}{m-i+1},
\end{align*}
reject all $H_{(i')}$ with $i'\leq i$. This procedure controls FWER under positive regression dependence on the subset of true null hypotheses (PRDS, \cite{benjamini2001control}), and is more powerful than Holm's procedure \citep{holm1979simple} in the sense that it never makes fewer rejections. 

For $m=2$ hypotheses, Hochberg's procedure has the same rejection region as the BH procedure for $R=0,1,2$, as depicted in the left panel of Figure~\ref{fig:MASL-Rej} (see also \cite{solari2017minimally}). In this case, SL never makes more rejections than Hochberg's procedure: if $(p_1,p_2)$ is in the triangle with vertices $(0,\alpha/2),(0,\alpha)$, and $(\alpha/2,\alpha)$ (or its symmetric counterpart below the diagonal), then the SL procedure makes one rejection whereas Hochberg's procedure makes two rejections. Outside of this region, the two procedures make the same number of rejections (see Figure~\ref{fig:MASL-Rej}). From this, and the established FWER control of Hochberg's procedure, it follows that bFDR of the SL procedure is controlled under PRDS when $m=2$.

\begin{proposition}
\label{prop:prds}
Let $m=2$ and let $(p_1,p_2)$ be PRDS on the subset of $p$-values corresponding to true null hypotheses, with $\P(p_i \leq t)\leq t$ for all $t \in [0,1]$ whenever $H_i$ is true. Then $\bFDR(\SL) \leq \alpha$ for every $\alpha \in (0,1)$.

\end{proposition}

\begin{proof}
For any multiple testing procedure, the boundary FDR is bounded by its FWER, since $\{H_{(R)} \text{ is true}, R>0\} \subseteq \{V \geq 1\}$, where $V$ is the number of false discoveries. Under PRDS, Hochberg's procedure controls $\FWER \leq \alpha$, since it is the closure of Simes' procedure when $m=2$. Therefore, since the SL procedure never makes more rejections than Hochberg's procedure,
\begin{align*}
    \bFDR(\SL) \leq \FWER(\SL) \leq \FWER(\text{Hochberg})\leq \alpha.
\end{align*}

\end{proof}

\subsection{Proofs for Table \ref{tab:procedure-error-control}}
\label{sec:proofs-table-1}

In this appendix section, we justify the entries in Table \ref{tab:procedure-error-control}, assuming throughout that the $m=2$ $p$-values are independent with Uniform$(0,1)$ null distribution.

For the SL procedure, FWER control follows from the proof of Proposition \ref{prop:prds}, bFDR control follows from Lemma 2 in \cite{soloff2024edge}, and FDR control follows from FWER control and the relation $\FDR \leq \FWER$. 

For the BH procedure, FWER control follows from its equivalence to Hochberg's procedure when $m=2$, as noted by \cite{solari2017minimally}. Boundary FDR control follows from the stated FWER control and the relation $\bFDR \leq \FWER$. We also note that the bFDR of the BH procedure can be higher than $\pi_0\alpha$. For example, take $\pi_0=1/2$ with $p_1 \sim \text{Uniform}(0,1)$ and $p_2 \equiv 0$. Then the boundary rejection is null when both hypotheses are rejected, which happens if and only if $p_1 < \alpha$. Therefore $\bFDR(\BH) = \alpha > \frac{1}{2}\alpha=\pi_0\alpha$.

For the MABH procedure, the FWER violation was noted by \cite{solari2017minimally}, who also proved FDR control. The bFDR violation for MABH was described in Section \ref{subsec:type-1-comparison}. 

For the MASL procedure, bFDR control is the conclusion of Corollary \ref{cor:MASL-bFDR-bound}. FDR control under the global null is equivalent to Simes' guarantee. Now suppose $m_0=1$ and $H_1$ is the true null. The FDR of MASL can be expressed as
\begin{align*}
    \FDR(\MASL) &= \P(R^+_\alpha=1, \; H_1\text{ is rejected}) + \frac{1}{2} \; \P(R^+_\alpha=2).
\end{align*}
Since $p_1$ and $p_2$ are independent, we may fix $p_2 \in [0,1]$ and analyze the above probabilities conditionally on the value of $p_2$. We split into two cases based on whether $p_2 \leq \alpha$. If $p_2 \leq \alpha$, then $H_1$ cannot be the sole rejection because this would require $p_1=p_{(1)}$ and
\begin{align*}
    p_1 - \alpha < p_2 - 2\alpha \Rightarrow p_2 > \alpha+p_1 > \alpha.
\end{align*}
If both hypotheses are rejected, then $p_{(2)} - 2\alpha < 0$, which implies $p_1 \leq 2\alpha$. Thus,
\begin{align*}
    \E\Big[ \frac{V}{1\vee R_\alpha^+} \mid p_2 \Big] &\leq \frac{1}{2} \cdot 2\alpha =\alpha.
\end{align*}
If $p_2 > \alpha$, then $R_\alpha^+ > 0$ requires $p_1=p_{(1)}<\alpha/2$ since $p_2>\alpha$. Hence $p_1 - \alpha/2 < 0$, so in this case we have
\begin{align*}
    \E\Big[ \frac{V}{1\vee R_\alpha^+} \mid p_2 \Big] &\leq \frac{\alpha}{2} \leq \alpha.
\end{align*}
Integrating over $p_2$ completes the proof of $\FDR(\MASL) \leq \alpha$ when $m=2$. 

To see why MASL can violate FWER control when $m=2$, consider $p_2\equiv \alpha/4$ fixed, with $p_1 \sim~\text{Uniform}(0,1)$ so that $m_0=1$ with $H_1$ being the true null. Since $p_{(1)} \leq \alpha/4 < \alpha/2$, we always have that $R_\alpha^+ > 0$. When $p_1\leq p_2$, both hypotheses are rejected if $p_2 - 2\alpha \leq p_1 - \alpha$, i.e.~$\alpha/4=p_2 \leq p_1 + \alpha$, which always holds. When $p_1 > p_2$, we have $p_2-\alpha = \alpha/4-\alpha<0$, so $H_2$ is always rejected. $H_1$ is rejected if and only if $p_1 -2\alpha < p_2-\alpha = -3\alpha/4$, i.e.~$p_1 < 2\alpha-3\alpha/4 = 5\alpha/4$. Putting both cases together, we find that $H_1$ is rejected if and only if $p_1 < 5\alpha/4$. Since $p_1\sim \text{Uniform}(0,1)$, this implies for $\alpha<4/5$,
\begin{align*}
    \FWER(\MASL) = 5\alpha/4 > \alpha.
\end{align*}

\noindent\textbf{Reproducibility.} Code to reproduce all figures is available on Github, at the repository: \url{https://github.com/sarahmostow/sarahmostow.github.io/tree/main/Minimally%20Adaptive%20Support%20Line}

\end{document}